\documentstyle[preprint,aps,12pt]{revtex}
\tightenlines
\title{The Contribution of Hot Electron Spin Polarization to the Magnetotransport in a Spin-Valve Transistor at Finite Temperatures}
\author{Jisang Hong and P. S. Anil Kumar}
\address{Max-Planck-Institut f{\"u}r Mikrostrukturphysik, Weinberg 2, D-06120 Halle, Germany}
\begin{document}
\maketitle
\begin{abstract}
The effect of spin mixing due to thermal spin waves and temperature dependence of hot electron spin polarization to the collector current in a spin-valve transistor has been theoretically explored. We calculate the collector current as well as the temperature dependence of magnetocurrent at finite temperatures to investigate the relative importance of spin mixing and hot electron spin polarization. In this study the inelastic scattering events in ferromagnetic layers have been taken into account to explore our interests. The theoretical calculations suggest that the temperature dependence of hot electron spin polarization has substantial contribution to the magnetotransport in the spin-valve transistor.
\end{abstract}
\newpage
\newcommand{\eb}{\begin{eqnarray}}
\newcommand{\ee}{\end{eqnarray}}
\section{Introduction}
Ultrathin magnetic multilayers exhibit unique properties not found in bulk materials. For example, magnetic tunneling junction (MTJ) \cite{junction} displays conductance strongly dependent on the relative orientation of the magnetization of the two ferromagnetic materials. Recently, spin-valve transistor \cite {valve} (SVT), as a new magnetoelectronic device, is suggested as well. This SVT has very different structure \cite{structure} and transport property from those of the conventional magnetic tunneling junction. In a SVT, electrons injected into the metallic base across a Schottky barrier (emitter side) penetrate the spin-valve base and reach the opposite side (collector side) of the transistor. When these injected electrons traverse the spin-valve they are above the Fermi level of the metallic base. Therefore, {\it hot} electron magnetotransport should be taken into account when one explores the collector current in a spin-valve transistor.  

When one discusses the transport of hot electrons in materials, one should note that the transport property of hot electrons is different from that of Fermi electrons. For instance, in magnetic tunneling junction (MTJ), spin polarization of Fermi electrons substantially depends on the density of states at the Fermi level. In contrast, the {\it hot} electron transport properties are related to the density of unoccupied states above the Fermi level, and it has an exponential dependence on inelastic mean free path \cite{mean}. Indeed, this electron inelastic mean free path plays a very important role when one explores transport property of hot electron. For example, Pappas {\it et al} \cite{Papps} measured substantial spin asymmetry in the electron transmission through ultrathin film of Fe deposited on Cu(100). This experiment implies that understanding of spin dependence of the inelastic mean free path is essential to the interpretation of the information obtained from spin polarized probes. In this regard, theoretical calculations of spin dependent inelastic electron mean free path \cite{free,path} have been presented. In these theoretical calculations, substantial spin dependent scattering rate was obtained. Along with this, spin dependence of scattering rate varying the energy of the probe beam electron was also explored. Generally, experimental data of the spin polarized electron spectroscopies  have been interpreted in terms of Stoner excitations when one discusses the spin dependent electron inelastic mean free path. Interestingly, the importance of spin wave excitations \cite{spin} in ferromagnetic Fe has been presented experimentally, and theoretical calculations \cite{path} also show that the spin wave excitations contribute significantly to the inelastic mean free path at low energies (roughly up to 1 eV above the Fermi level). These experimental and theoretical evidences suggest that the property of spin wave excitations at low energy should be investigated more in detail in relation to the spin-valve transistor as the electron energy in a SVT is roughly 1 eV above the Fermi level.   

In the spin-valve transistor structure, Jansen {\it et al} reported very interesting behaviors of the spin dependent collector current at finite temperatures \cite{finite}. The measured collector current across the spin-valve shows unusual features depending on the relative orientation of the magnetic moment in the ferromagnetic layers at finite temperatures. When the magnetic moments are parallel in each layer the collector current (parallel collector current) is increasing up to 200 K and decreasing beyond that temperature regime, while the anti-parallel collector current is increasing up to room temperature. Generally speaking, the scattering strength increases with temperature T in ordinary metals. This implies that any thermally induced scattering process enhances the total scattering. One then expects that the observed collector current will be decreasing with increasing temperature T in any spin configuration of the spin-valve transistor. Hence, the temperature dependence of measured collector current may not be related to the ordinary scattering events in the metallic base. Two different processes are suggested to explain this observation in the spin-valve transistor. One is the spatial distribution of Schottky barrier. With increasing temperature T electrons have more chances to overcome the Schottky barrier at the collector side because of the shift of the injected electron energy due to thermal energy. This mechanism, however, can not account for the temperature dependence of collector current beyond 200 K. Besides, Schottky barrier distribution does not have any spin dependent property, but only have the influence on the magnitude of both the parallel and anti-parallel collector current. Thus, authors of Ref. \cite{finite} attribute measured temperature dependence of the collector current to the spin mixing effect due to thermal spin waves, and this is supported by the temperature dependence of magnetocurrent. 

It will be of interest to estimate the effect of spin mixing doe to thermal spin waves to the collector current at finite temperatures. Along with this spin mixing mechanism, we believe that {\it hot} electron spin polarization may also have an influence on the collector current. Unfortunately, the hot electron spin polarization at low energy (roughly speaking 1 eV above the Fermi level) has not been extensively explored. Although there is an example of lifetime measurement of Co \cite{Co} in the relevant energy regime to the spin-valve transistor, it does not contain any  data about the temperature dependence.  In this calculations we therefore model the hot electron spin polarization. This will be discussed below. Now, the main interest in our work is in understanding what gives the substantial contribution to the behaviors both of the parallel and anti-parallel collector current at finite temperatures.  Magnetocrrent may not be a useful quantity for our purpose because magnetocurrent depends on the difference of the parallel and anti-parallel collector current. In addition, it is also influenced by the magnitude of the anti-parallel collector current by the definition of magnetocurrent \cite{finite}. Therefore, even a small change in parallel and anti-parallel collector current may affect dramatically to the magnetocurrent.  We then have to calculate both the parallel and anti-parallel collector current to explore the issue raised in Ref. \cite{finite}. 

If one is interested in the absolute magnitude of the collector current one obviously needs to take into account many spin independent scattering events as well as spin dependent scattering processes. Along with this one may also consider angle dependence \cite{angle} even if electrons have enough energy to overcome the Schottky barrier at the collector side. In addition, Schottky barrier height distribution \cite{structure} can also affect the magnitude of collector current. Our interest, once again, is in understanding the effect of spin mixing and hot electron spin polarization to the parallel and anti-parallel collector at finite temperatures. In the spirit of this, we do not include inelastic scattering effect in normal metal layers. An exponential dependence on the inelastic mean free path of the (both parallel and anti-parallel) collector current \cite{mean} enables us to consider only the events in ferromagnetic layers when we focus our interests on the issue raised in the Ref. \cite{finite}.   

\section{Model}
A spin-valve transistor has typically $Si/N/F/N/F/N/Si$ structure \cite{structure} where $N$ denotes normal metal, and $F$ represents ferromagnetic metal. Since the injected electrons into the spin-valve across the Schottky barrier at the emitter side are not spin polarized before they pass through the first ferromagnetic layer the number of spin up and spin down electrons will be the same. Thus, we can say that $N_0$ spin up and spin down electrons pass across the Schottky barrier per unit time per unit area, respectively. When these electrons penetrate ferromagnetic layer some of injected electrons will be lost due to inelastic scattering events in the spin-valve base. We define $\gamma_{M(m)}(T)$ to describe the inelastic scattering for majority (minority) spin electrons in ferromagnetic material at temperature T. We can relate this $\gamma_{M(m)}(T)$ to the inelastic mean free path such as $\gamma_{M(m)}(T)=exp[-w/l_{M(m)}(T)]$ where $l_{M(m)}(T)$ is the inelastic mean free path of majority (minority) spin electron in ferromagnetic material at temperature T, and $w$ is the thickness of that material. For example, with initial $N_0$ injected electrons $N_0\gamma_{M(m)}(T)$ electrons will pass the ferromagnetic layer if they are majority (minority) spin electrons. This $\gamma_{M(m)}(T)$ is related to the spin polarization of hot electrons. One should note that spin polarization of hot electrons enters into the spin-valve system, not that of Fermi electrons. 

Hot electron spin polarization $P_H(T)$ and spin flip probability $P(T)$ are essential quantities to explore our main issue. This spin flip probability due to thermal spin wave emission or absorption gives spin mixing effect. It has been shown in Ref. \cite{free} that scattering rate of both majority and minority spin electron resulting from thermal spin wave emission and absorption is virtually the same at low temperatures, then the spin flip probability from thermal spin wave emission and absorption will have the same temperature dependence. If spin-flip process is operating, the parallel collector current from spin-up source electrons can be calculated in the following way. $N_0\gamma_M(T)$ electrons penetrate the first ferromagnetic metal layer. Among these electrons, $N_0\gamma_M(T)(1-P(T))$ electrons keep their spin-up state, and $N_0\gamma_M(T)(1-P(T))\gamma_M(T)$ electrons will be collected with spin-up state. Along with that, $N_0\gamma_M(T)P(T)$ electrons are created having the opposite spin state resulting from spin-flip process as well, and $N_0\gamma_M(T)P(T)\gamma_m$ electrons are collected with the spin-down. Finally, the total number of collected electrons from spin-up electrons become $N_0\{\gamma_M^2(T)(1-P(T))+\gamma_M(T)\gamma_m(T)P(T)\}$. One can follow the same scheme to calculate the contribution to the current from spin-down source electrons, and also readily obtain the expression of anti-parallel collector current. We then write the parallel collector current as  
\eb
\tilde{I}_c^P(T,P(T))=N_0\gamma_M^2(T)[\{1+(\frac{\gamma_m(T)}{\gamma_M(T)})^2\}(1-P(T))
+2(\frac{\gamma_m(T)}{\gamma_M(T)})P(T)]. 
\ee
Similarly, the anti-parallel collector current becomes
\eb
\tilde{I}_c^{AP}(T,P(T))=N_0\gamma_M^2(T)[\{1+(\frac{\gamma_m(T)}{\gamma_M(T)})^2\}P(T)
+2(\frac{\gamma_m(T)}{\gamma_M(T)})(1-P(T))]. 
\ee
As remarked earlier we can relate the $\gamma_M(T)$ and $\gamma_m(T)$ to the hot electron spin polarization $P_H(T)$ at finite temperatures.  We write this
\eb
\frac{\gamma_m(T)}{\gamma_M(T)}=\frac{1-P_H(T)}{1+P_H(T)}.
\ee 
From this expression, most generally we express the $\gamma_M(T)$ and $\gamma_m(T)$ as
\eb
\gamma_M(T)=g(T)(1+P_H(T))
\ee
and
\eb
\gamma_m(T)=g(T)(1-P_H(T))
\ee
where $g(T)$ is a function of temperature T. This function $g(T)$ enters into the both $\gamma_M(T)$ and $\gamma_m(T)$ simultaneously so that the detailed form of $g(T)$ does not affect our purpose of this work, save for the magnitude both of the parallel and anti-parallel collector current expressed in Eqs. (1) and (2). Therefore, when one explores the effect of spin mixing and temperature dependence of hot electron spin polarization to the collector current depending on the relative spin orientation of the ferromagnetic layers one can treat the function $g(T)$ simply as a prefactor. This enables us to explore the collector current, expressed below, to study the our interest in this work. By substitution the Eqs. (4) and (5) into Eqs. (1) and (2), we obtain 
\eb
I_c^P(T,P(T))=N_0(1+P_H(T))^2 [\{1+(\frac{1-P_H(T)}{1+P_H(T)})^2\}(1-P(T))+2\frac{1-P_H(T)}{1+P_H(T)}P(T)]
\ee
and
\eb
I_c^{AP}(T,P(T))=N_0(1+P_H(T))^2 [\{1+(\frac{1-P_H(T)}{1+P_H(T)})^2\}P(T)+2\frac{1-P_H(T)}{1+P_H(T)}(1-P(T))].
\ee
In these equations, we do not take into account the effect of $g(T)$ as explained above. One can also understand that magnetocurrent satisfies the following property by the definition of magnetocurrent
\eb
MC(T,P(T))\equiv\frac{I_c^P(T,P(T))-I_c^{AP}(T,P(T))}{I_c^{AP}(T,P(T))}=\frac{\tilde{I}_c^P(T,P(T))-\tilde{I}_c^{AP}(T,P(T))}{\tilde{I}_c^{AP}(T,P(T))}.
\ee
As we mentioned, temperature dependence of hot electron spin polarization at low energy regime (roughly speaking, 1 eV above the Fermi level) has not been investigated actively. In our calculations, we test two cases such as 
\eb
P_H(T)=P_0(1-[\frac{T}{T_c}]^{3/2})
\ee
and
\eb
P_H(T)=P_0(1-[\frac{T}{T_c}])
\ee
where $P_0$ is the hot electron spin polarization at zero temperature, and $T_c$ is the critical temperature of ferromagnetic metal of interest. In our calculations we take $T_c=650 K$ to simulate pseudo permalloy. We also assume that spin mixing probability $P(T)$ has $T^{3/2}$ dependence with temperature T by the virtue of the fact that number of thermal spin waves \cite{free} are proportional to $T^{3/2}$. We then write this as $P(T)=cT^{3/2}$ where $c$ is a parameter. In this calculations we limit the temperature ranges from zero to room temperature (T=300 K) as reported in the experiment \cite{finite}. If we define $P_r$ as a spin flip probability at room temperature (300 K), the parameter $c$ in $P(T)$ can be written as $c=P_r \times [\frac{1}{300 K}]^{3/2}$. One then can express the $P(T)$ as $P(T)=P_r \times [\frac{T}{300K}]^{3/2}$. We take the maximum spin flip probability at 300 K in our numerical calculations to explore our interests. (The maximum spin flip probability is 0.5, and  one can understand this from Eqs. (1) and (2).)

\section{Results and Discussions}
Now, we discuss the results of our model calculations. Fig. 1(a) and (b) display the collector current expressed in Eqs. (6) and (7) without spin mixing effect (here, $P(T)=0$ and $P_H(T)=P_0(1-[\frac{T}{T_c}]^{3/2})$. The parallel collector current is normalized at T=0, and the anti-parallel collector current is the relative magnitude with respect to the parallel collector current. One can clearly see that the parallel collector current is decreasing, and the anti-parallel collector current is increasing with temperature T. The $1+P_H(T)$ and $1-P_H(T)$ behave the opposite way with temperature T. Thus, those two terms are competing each other and contributing differently to the parallel and anti-parallel collector current. Fig. 2(a) and (b) represent the collector current in Eqs. (6) and (7) including the spin mixing effect with the same $P_H(T)$ as in Fig. 1. One can see that the parallel and anti-parallel collector have been changed after including spin mixing. However, the deviation from the results in Fig. 1 is not substantial. To evaluate  how much the collector current is influenced by introducing the spin mixing effect, we calculate the quantity $R(T)=[I_c^P(T,0)-I_c^P(T,P(T))]/I_c^P(T,0)$. Fig. 3(a) presents the $R(T)$ with $P_H(T)=P_0(1-[\frac{T}{T_c}]^{3/2})$, and Fig. 3(b) is the case with $P_H(T)=P_0(1-[\frac{T}{T_c}])$. From the Fig. 3(a) and 3(b), We find that the parallel collector current has been changed roughly 10 \% when $P_0=0.6$. For anti-parallel collector, we obtain almost the same result. This implies that a substantial temperature dependence of collector current depending on the relative spin orientation in ferromagnetic layers can be explained by only taking into account the hot electron spin polarization, without introducing spin mixing mechanism. We interprete our results as following. We estimate the wave vector of thermally excited spin waves by setting $DQ_T^2$ equal to $k_BT$, then at room temperature  $Q_T \approx 0.3 A^{-1}$ if we take $D\approx 400$ meV-$A^2$. This is a small fraction of the distance to the zone boundary. Therefore, only a few percent of the Brillouin zone contains thermally excited spin waves. One should note that $T^{3/2}$ dependence of thermal spin waves is obtained we integrate over the whole Brillouin zone. However, only a small volume of the Brillouin zone contributes to the thermal spin waves at room temperature we therefore have even weaker temperature dependence of thermal spin waves than we have modeled in our work. As a result, spin mixing mechanism marginally contributes to the both parallel and anti-parallel collector current at finite temperatures. In Fig.4 , we present the magnetocurrent without spin mixing effect. This is also normalized at T=0. One can see that the magnetocurent also accords with experimental data of Ref. \cite {finite} semi-quantitatively. 

In conclusion, we have explored the collector current in a spin-valve transistor at finite temperatures to understand the influence of spin mixing due to thermal spin waves and temperature dependence of hot electron spin polarization on the collector current. We obtain that hot electron spin polarization contributes substantially to the collector current at finite temperatures compared to the effect of spin mixing. Here, we do not claim that there is no spin mixing mechanism due to thermal spin waves. Once again, when we discuss the relative importance of spin mixing and hot electron spin polarization effect to the collector current at finite temperatures we suggest that major temperature dependence of collector current stems from temperature dependence of hot electron spin polarization even if we have spin mixing process. We hope that this work will stimulate further related studies such as hot electron spin polarization at finite temperatures and temperature dependence of electron inelastic mean free path at low energy. 

\newpage
\begin{figure}
\caption{(a) The parallel collector current $I_C^P(T,0)$ expressed in Eq. (6) with normalization at T=0. Here, we take $P_H(T)$ as $P_H(T)=P_0(1-[\frac{T}{T_c}]^{3/2})$. (b) The anti-parallel collector current $I_C^{AP}(T,0)$ expressed in Eq. (7). This is the relative magnitude with respect to the parallel collector current. The same hot electron spin polarization is taken into account.}
\end{figure}
\begin{figure}
\caption{(a) The parallel collector current $I_c^P(T,P(T))$ expressed in Eq. (6) with normalization at T=0 including spin mixing effect. $P_H(T)=P_0(1-[\frac{T}{T_c}]^{3/2})$ is used in this calculation. (b) The anti-parallel collector current $I_c^{AP}(T,P(T))$ expressed in Eq. (7). This is the relative magnitude with respect to the parallel collector current. The same hot electron spin polarization is used. }
\end{figure}
\begin{figure}
\caption{(a) The ratio of $[I_c^P(T,0)-I_c^P(T,P(T))]/I_c^P(T,0)$ with $P_H(T)=P_0(1-[\frac{T}{T_c}]^{3/2})$ (b) The same quantity as in (a) with $P_H(T)=P_0(1-[\frac{T}{T_c}])$ }
\end{figure}
\end{document}